\documentclass[12pt]{article}
\usepackage{amsfonts}
\usepackage{amssymb}
\usepackage{amsmath}

\def\hybrid{\topmargin -20pt    \oddsidemargin 0pt
        \headheight 0pt \headsep 0pt
        \textwidth 6.25in       
        \textheight 9.25in       
        \marginparwidth .875in
        \parskip 5pt plus 1pt   \jot = 1.5ex}

\hybrid

\def\baselinestretch{1.2}

\catcode`\@=11

\def\marginnote#1{}
\newcount\hour
\newcount\minute
\newtoks\amorpm
\hour=\time\divide\hour by60
\minute=\time{\multiply\hour by60 \global\advance\minute by-\hour}
\edef\standardtime{{\ifnum\hour<12 \global\amorpm={am}%
        \else\global\amorpm={pm}\advance\hour by-12 \fi
        \ifnum\hour=0 \hour=12 \fi
        \number\hour:\ifnum\minute<10 0\fi\number\minute\the\amorpm}}
\edef\militarytime{\number\hour:\ifnum\minute<10 0\fi\number\minute}

\def\draftlabel#1{{\@bsphack\if@filesw {\let\thepage\relax
   \xdef\@gtempa{\write\@auxout{\string
      \newlabel{#1}{{\@currentlabel}{\thepage}}}}}\@gtempa
   \if@nobreak \ifvmode\nobreak\fi\fi\fi\@esphack}
        \gdef\@eqnlabel{#1}}
\def\@eqnlabel{}
\def\@vacuum{}
\def\draftmarginnote#1{\marginpar{\raggedright\scriptsize\tt#1}}

\def\draft{\oddsidemargin -.5truein
        \def\@oddfoot{\sl preliminary draft \hfil
        \rm\thepage\hfil\sl\today\quad\militarytime}
        \let\@evenfoot\@oddfoot \overfullrule 3pt
        \let\label=\draftlabel
        \let\marginnote=\draftmarginnote
   \def\@eqnnum{(\theequation)\rlap{\kern\marginparsep\tt\@eqnlabel}%
\global\let\@eqnlabel\@vacuum}  }

\def\preprint{\twocolumn\sloppy\flushbottom\parindent 2em
        \leftmargini 2em\leftmarginv .5em\leftmarginvi .5em
        \oddsidemargin -.5in    \evensidemargin -.5in
        \columnsep .4in \footheight 0pt
        \textwidth 10.in        \topmargin  -.4in
        \headheight 12pt \topskip .4in
        \textheight 6.9in \footskip 0pt
        \def\@oddhead{\thepage\hfil\addtocounter{page}{1}\thepage}
        \let\@evenhead\@oddhead \def\@oddfoot{} \def\@evenfoot{} }

\def\numberbysection{\@addtoreset{equation}{section}
        \def\theequation{\thesection.\arabic{equation}}}

\def\underline#1{\relax\ifmmode\@@underline#1\else
        $\@@underline{\hbox{#1}}$\relax\fi}

\def\titlepage{\@restonecolfalse\if@twocolumn\@restonecoltrue\onecolumn
     \else \newpage \fi \thispagestyle{empty}\c@page\z@
        \def\thefootnote{\fnsymbol{footnote}} }

\def\endtitlepage{\if@restonecol\twocolumn \else \newpage \fi
        \def\thefootnote{\arabic{footnote}}
        \setcounter{footnote}{0}}  

\catcode`@=12
\relax

\def\figcap{\section*{Figure Captions\markboth
        {FIGURECAPTIONS}{FIGURECAPTIONS}}\list
        {Figure \arabic{enumi}:\hfill}{\settowidth\labelwidth{Figure
999:}
        \leftmargin\labelwidth
        \advance\leftmargin\labelsep\usecounter{enumi}}}
 \relax
\def\tablecap{\section*{Table Captions\markboth
        {TABLECAPTIONS}{TABLECAPTIONS}}\list
        {Table \arabic{enumi}:\hfill}{\settowidth\labelwidth{Table
999:}
        \leftmargin\labelwidth
        \advance\leftmargin\labelsep\usecounter{enumi}}}
 \relax
\def\reflist{\section*{References\markboth
        {REFLIST}{REFLIST}}\list
        {[\arabic{enumi}]\hfill}{\settowidth\labelwidth{[999]}
        \leftmargin\labelwidth
        \advance\leftmargin\labelsep\usecounter{enumi}}}
 \relax
\makeatletter
\newcounter{pubctr}
\def\publist{\@ifnextchar[{\@publist}{\@@publist}}
\def\@publist[#1]{\list
        {[\arabic{pubctr}]\hfill}{\settowidth\labelwidth{[999]}
        \leftmargin\labelwidth
        \advance\leftmargin\labelsep
        \@nmbrlisttrue\def\@listctr{pubctr}
        \setcounter{pubctr}{#1}\addtocounter{pubctr}{-1}}}
\def\@@publist{\list
        {[\arabic{pubctr}]\hfill}{\settowidth\labelwidth{[999]}
        \leftmargin\labelwidth
        \advance\leftmargin\labelsep
        \@nmbrlisttrue\def\@listctr{pubctr}}}
 \relax
\makeatother
\newskip\humongous \humongous=0pt plus 1000pt minus 1000pt

\newif\ifdtup

\relax

\def\be{\begin{equation}}
\def\ee{\end{equation}}
\def\ba{\begin{eqnarray}}
\def\ea{\end{eqnarray}}

\def\del{\partial}

\def\a{\alpha}

\def\b{\beta}

\def\g{\gamma}
\def\G{\Gamma}
\def\d{\delta}
\def\D{\Delta}
\def\e{\epsilon}

\def\P{\Pi}

\def\th{\theta}
\def\Th{\Theta}
\def\m{\mu}
\def\n{\nu}
\def\om{\omega}
\def\Om{\Omega}
\def\l{\lambda}
\def\L{\Lambda}
\def\s{\sigma}
\def\S{\Sigma}

\def\cL{{\cal L}}

\def\no{\noindent}

\def\qq{\qquad}

\def\IR{\relax{\rm I\kern-.18em R}}

\def \ha {{1\over 2}}

\def \ov {\over}

\def\II{\relax{\rm 1\kern-.35em1}}
\def\IR{\relax{\rm I\kern-.18em R}}
\def\inv{^{\raise.15ex\hbox{${\scriptscriptstyle -}$}\kern-.05em 1}}

\def\cL{{\cal L}}

\begin{document}
\renewcommand{\theequation}{\arabic{equation}}

\newcommand{\beq}{\begin{equation}}
\newcommand{\eeq}[1]{\label{#1}\end{equation}}
\newcommand{\ber}{\begin{eqnarray}}
\newcommand{\eer}[1]{\label{#1}\end{eqnarray}}
\newcommand{\eqn}[1]{(\ref{#1})}
\begin{titlepage}
\begin{center}

\hfill CPHT-RR052.0606 \vskip -.3 cm
\hfill NEIP--06--06 \\
\hfill hep--th/0610055\\

\vskip  0.5in

{\large \bf Non-Abelian coset string backgrounds from \\
asymptotic and initial data}

\vskip 0.4in

{\bf P. Marios Petropoulos$^{1}$}\phantom{x} and\phantom{x} {\bf
Konstadinos Sfetsos}$^{2}$ \vskip 0.1in

${}^1\!$ Centre de Physique Th{\'e}orique, Ecole Polytechnique${}^\dagger\!$\\
91128 Palaiseau Cedex, France\\ and \\
${}^{}\!$ Institut de Physique, Universit\'e de Neuch\^atel\\
Breguet 1, 2000 Neuch\^atel, Switzerland\\
{\footnotesize{\tt marios@cpht.polytechnique.fr}}

\vskip .2in

${}^2\!$
Department of Engineering Sciences, University of Patras\\
26110 Patras, Greece\\
{\footnotesize{\tt sfetsos@upatras.gr}}\\

\end{center}

\vskip .3in

\centerline{\bf Abstract}

\no
We describe hierarchies of exact string backgrounds obtained
as non-Abelian cosets of orthogonal groups and having a
space--time realization in terms of gauged WZW models. For each
member in these hierarchies, the target-space backgrounds are
generated by the ``boundary'' backgrounds of the next member. We
explicitly demonstrate that this property holds to all orders in
$\alpha'$. It is a consequence of the existence of an integrable
marginal operator build on, generically, non-Abelian parafermion
bilinears. These are dressed with the dilaton supported by the
extra radial dimension, whose asymptotic value defines the
boundary. Depending on the hierarchy, this boundary can be
time-like or space-like with, in the latter case, potential cosmological
applications.

\vfill
\hrule width 6.7cm \vskip.1mm{\small \small \small \noindent
$^\dagger$\  Unit{\'e} mixte  du CNRS et de  l'Ecole
Polytechnique, UMR 7644.}
\end{titlepage}
\vfill
\newpage
\setcounter{footnote}{0}
\renewcommand{\thefootnote}{\arabic{footnote}}


\setcounter{section}{0}

\def\baselinestretch{1.2}
\baselineskip 20 pt
\noindent


\tableofcontents

\section{Introduction}

In the string literature there has been a class of exact models describing consistent
string propagation based on $G/H$ conformal field theory (CFT) coset models \cite{coset}.
These
have a space--time realization in terms of gauged Wess--Zumino--Witten (WZW) models \cite{gwzw}.
In the gauged WZW models the action of the subgroup is generically vectorial and hence
fundamentally different than in geometric cosets.
Therefore, in order to distinguish them from geometric cosets we will call
them henceforth \emph{conformal cosets}.
When a certain procedure is followed, a metric, a dilaton and an
antisymmetric tensor field arise and the
conditions for conformal invariance are satisfied.
In the present work we revisit certain conformal coset
models based on orthogonal groups.

\no In particular, we will consider the Euclidean-signature
conformal cosets \be \mathrm{C}H_{d,k} = {SO(d,1)_{-k}\ov
SO(d)_{-k}}\ ,\qq d=2,3,\dots \label{ch} \ee and \be
\mathrm{C}S_{d,k} = {SO(d+1)_k\ov SO(d)_k}\ , \qq d=2,3,\dots
\label{cs1} \ee In addition, we will consider the corresponding
Minkowskian-signature ones \be \mathrm{CAdS}_{d,k} =
{SO(d-1,2)_{-k}\ov SO(d-1,1)_{-k}}\ ,\qq d=2,3,\dots \label{ch8}
\ee and \be \mathrm{CdS}_{d,k} = {SO(d,1)_k\ov SO(d-1,1)_k}\ ,\qq
d=2,3,\dots \label{cs8} \ee The indices $k$ and $-k$ indicate the
level of the corresponding current algebras and will some times be
omitted in order to simplify the notation, if there is no risk of
confusion. We have given them names reminiscent of the
corresponding geometric cosets (spheres, anti-de or de Sitter and
hyperbolic planes), but we should keep in mind that \emph{they are
different in various respects}.\footnote{In geometric cosets, the
action of the subgroup is one-sided. Considered as target spaces
of sigma-models, geometric cosets lead in general to non-vanishing
$\b$-functions and cannot therefore describe consistent string
propagation. The exceptions include the three-dimensional anti-de
Sitter and the three-sphere, because these are also group
manifolds, and cases where, dividing by a subgroup of the Cartan
torus, conformal invariance is restored by switching on $U(1)$
background gauge fields \cite{IKOP2, LS04}. Combinations of
maximally symmetric geometric cosets (anti-de Sitter spaces or
spheres), as those emerging in near-horizon geometries of brane
distributions, can appear as supergravity solutions. We also note the
work in \cite{LustCoset} where adding torsion to some
six-dimensional non-maximally symmetric geometric cosets, made
possible to retain conformal invariance.} In order to avoid a
confusion between geometric and conformal cosets, we have added
the letter ``C'' in the name of the latter, besides the level
indices $k$ or $-k$.

\no
The levels in \eqn{ch} and \eqn{ch8} are $-k$ for
ensuring that only a single time coordinate
appears. The class of models in \eqn{ch8} was suggested and analyzed in \cite{BN}
from an algebraic view point
using non-compact current algebras.
For the lower-dimensional cases
the explicit forms of the corresponding string theory backgrounds
have been explicitly worked out to lowest order in $\a'\sim 1/k$, for $d=2$
in \cite{wittenbh}
and for $d=3,4$ in \cite{BS1,BS2,BS3}.\footnote{The Euclidean
version of the two-dimensional solution was discussed
in \cite{BCR} in the context of classical Abelian parafermions and also found in
\cite{EFR,MSW} from the beta-functions point of view.}
Moreover, all perturbative $\a'$-corrections can be systematically worked out
using a combination of algebraic CFT and space--time techniques
developed in full generality in \cite{BSexa}.
Explicit results are worked out for the case with $d=2$ in \cite{BSexa,DVV}
and for $d=3,4$ in \cite{BSexa}.

\no
In this paper we will mainly focus on the cosets \eqn{ch}--\eqn{cs8}
aiming at uncovering their possible geometrical and CFT relations.
We will perform most of the computational details for the backgrounds
corresponding to the
Euclidean-signature conformal cosets.
The conclusions we will draw
for the Minkowskian-signature ones are easily
derived along essentially the same lines.
We claim that the spatial infinity of the space for the non-compact coset
$\mathrm{C}H_{d}$ is the space for the compact coset
$\mathrm{C}S_{d-1}$ times the linear dilaton $\mathbb{R}_Q$, where $Q$ is
an appropriate background charge. This will be exhaustively argued for
in Sec. 2, where
we will present the general structure of the models and of our method and in addition
for the benefit of the reader in the subsequent discussion, the general conclusions concerning all
possible relations involving the conformal cosets \eqn{ch}--\eqn{cs8}.

\no
In Secs. 3 and 4 we will further support the above claim by
explicit computations for the lowest
dimensional $d=3$ and $d=4$ cosets, respectively, which are already non-trivial,
especially the four-dimensional one.
The computation will involve the detailed comparison of the background fields
at the semiclassical as well as exact in $\a'$ levels.
In Sec. 5 we will go further and show that
the full coset $\mathrm{C}H_{d}$ can be thought of as arising via an exactly
marginal perturbation involving a bilinear in the chiral and
antichiral parafermions of the compact-coset theory $\mathrm{C}S_{d-1}$
dressed appropriately with a vertex operator in the linear dilaton
theory $\mathbb{R}_Q$. The anomalous dimension of the parafermions is precisely cancelled out
by that of the vertex operator, so that the perturbing operator has
dimension $(1,1)$ exactly. This property gives another perspective to the
observation that the asymptotic region of the coset $\mathrm{C}H_{d}$ is the
coset $\mathrm{C}S_{d-1}$ plus a free field: these two sigma models appear as
the end points of a continuous line of exact theories sharing all the same
asymptotic properties.

\section{General structure of the background fields}

The large-$k$ regime background fields for the conformal coset models \eqn{ch}--\eqn{cs8}
follows
from the corresponding gauged WZW action after a unitary gauge
that fixes a number a parameters equal
to the dimension of the subgroup $H$ is chosen and the corresponding gauge fields are integrated out.
In general, there is only a metric and a dilaton field,
whereas the antisymmetric tensor is zero, irrespective of the dimension $d$.
By construction the one-loop $\b$-function equations are satisfied.
Due to the complexity of the procedure, explicit results are known for the
semiclassical large-$k$ regime as well as exactly in $\a'\sim 1/k$, only for the
lower-dimensional cases $d=2,3$ and $4$.

\boldmath
\subsection{Large-$k$ semiclassical regime}
\unboldmath

From the unifying treatment of \cite{BS3,BSexa} as well as experience with the
low-dimensional cases, we deduce that the structure of the
metric of the general coset models in \eqn{ch}--\eqn{cs8} in the semiclassical large-$k$-regime is
\ba
\mathrm{d}s^2_d = 2 k \left[  {\mathrm{d}b^2\ov 4 (b^2-1)} +
g_{ij}(b,{\bf x})\,  \mathrm{d}x^i\, \mathrm{d}x^j \right] \ ,
\qq i = 1,2,\dots, d-1\ ,
\label{smk1}
\ea
whereas for the dilaton, the corresponding expression is
\ba
 \mathrm{e}^{-2\Phi}= \mathrm{e}^{-2\Phi_0} (b+1)(b-1)^{d-2} F({\bf x})\ .
\label{fh4}
\ea
For the case of the $\mathrm{C}H_{d,k}$ coset in which we work out all details for $d=3$ and $d=4$
in Secs. 3 and 4 below, the variable $b$ is non-compact and takes values
in the open intervals with $|b|>1$.
The set of variables $x^i$ can be chosen such that they are all compact
and take values in finite intervals. The metric components $g_{ij}(b,{\bf x})$
and the function $F({\bf x})$ have specific forms depending on the particular model.
In the limit $b\to \infty$ this variable decouples and corresponds to a free
boson. In particular, letting
\ba
b= \mathrm{e}^{2 x}  \ ,
\label{bx}
\ea
in the limit $x\to \infty$ the metric takes the form\footnote{An alternative
equivalent way of viewing this limit is to first rescale the variable $b$ as
$b\to b/\l$ and then take the limit $\l\to 0$. In this way the background
\eqn{smk1} and \eqn{fh4} can be thought of as
the integrated perturbation of the leading order correction to the $\l\to 0$
result.}
\ba
\mathrm{d}s^2_d = 2 k \left( \mathrm{d}x^2 + \hat g_{ij}({\bf x})\, \mathrm{d}x^i \, \mathrm{d}x^j \right)\ ,
\label{h3}
\ea
where the relation to the metric components $g_{i,j}(b,{\bf x})$ in \eqn{smk1} is
\ba
\qq \hat g_{ij}({\bf x}) = g_{ij}(\infty,{\bf x})\ .
\label{h5}
\ea
The case with $b\to -\infty$ is equivalent to that
with $b\to \infty$, since we may instead of
\eqn{bx} define $b=- \mathrm{e}^{2 x}$, then let $x\to \infty$ and note that in our models it turns out that
$g_{ij}(-\infty,{\bf x})= g_{ij}(\infty,{\bf x})$.
Similarly for the dilaton
\ba
 \mathrm{e}^{-2\Phi}=  \mathrm{e}^{-2 \Phi_0}  \mathrm{e}^{2(d-1)x } F({\bf x})\ ,
 \qq {\rm as}\quad x\to \infty\ .
\label{j8}
\ea
Since in this limit for generic values of ${\bf x}$ we have that
$ \mathrm{e}^\Phi\to 0$, the large-$b$ limit is a weak-string coupling limit.
We claim that the decoupled $(d-1)$-dimensional metric and the corresponding dilaton
\ba
\mathrm{d}s_{(d-1)}^2= 2 k \hat g_{ij}({\bf x})\, \mathrm{d}x^i\, \mathrm{d}x^j\ ,
\qq  \mathrm{e}^{-2\Phi_{(d-1)}}= F({\bf x})\ ,
\label{fdk}
\ea
correspond to the semiclassical large-$k$ regime background for the compact coset model
$\mathrm{C}S_{d-1,k}$.
We will explicitly demonstrate these in the lowest-dimensional cases with $d=3,4$
which already are highly non-trivial, especially the four-dimensional one.
This demonstration is not as straightforward as it might seem up to now since
it will involve finding the necessary, quite complicated,
coordinate transformations that will make the
metrics look identical.
The Liouville field contributes asymptotically linearly only to the dilaton
term as $-(d-1)x$.

\boldmath
\subsection{All-$k$ exactness}
\unboldmath

In \cite{BSexa} a method to algebraically deduce
all $\a'\sim 1/k$ perturbative corrections to
the low-energy metric and dilaton was found for any conformal coset.
Using this as well as the explicit results for the
low-dimensional cases of the general coset models in \eqn{ch}--\eqn{cs8}
we deduce that the form of the exact metric is
\ba
\mathrm{d}s^2 = 2(k-d+1)\left[{\mathrm{d}b^2\ov 4(b^2-1)} +
g_{ij}(b,{\bf x};k)\, \mathrm{d}x^i\, \mathrm{d}x^j \right]\ ,\qq
i =1,2,\dots, d-1\ ,
\label{fl1}
\ea
where we have indicated that the $(d-1)$-dimensional metric $g_{ij}$ depends
explicitly on the level $k$.
Similarly, for the dilaton we deduce that the exact expression is of the form
\ba
 \mathrm{e}^{-2\Phi}= \mathrm{e}^{-2\Phi_0} {1\ov \sqrt{\b(b,{\bf x};k)}} \ (b+1)(b-1)^{d-2} F({\bf x})\ ,
\label{fh34}
\ea
where the function $\b$ above is such that $\b(b,{\bf x};\infty)=1$.
Using that and the relation to the semiclassical
$(d-1)$-dimensional
metric appearing in \eqn{smk1}
$g_{ij}(b,{\bf x};\infty)= g_{ij}(b,{\bf x})$, we see that the exact background fields
\eqn{fl1} and \eqn{fh34} go smoothly over to the semiclassical counterparts in
\eqn{smk1} and \eqn{fh4}.
We also note that since the measure given by the combination
\ba
 \mathrm{e}^{-2\Phi }\sqrt{\det(G_{\m\n})} \ ,
\label{meas}
\ea
is $k$-independent \cite{BSexa} we have the relation
\ba
\det(g_{ij}(b,{\bf x};k)) = \b(b,{\bf x};k)\ \det(g_{ij}(b,{\bf x})) \ .
\label{mkl2}
\ea

\no
We claim that, in the limit $b\to \infty$, the exact matching involves also a
shift of the level $k$ as $k\to k+2 d -4$. These will be shown by
comparing the exact $\a'$-corrected geometries for the three- and four-dimensional
cases in Secs. 3 and 4, as well as when we examine the nature of the marginal
perturbation, involving dressed parafermions in Sec. 5, in any dimension.
For the time being let's assume that and
recall that the central charges corresponding
to the models \eqn{ch} and \eqn{cs1} are
\ba
c_{\mathrm{C}H}(d,k)={d(d+1)k\ov 2 (k-d+1)}-{d(d-1)k\ov 2 (k-d+2)}\ ,
\nonumber\\
c_{\mathrm{C}S}(d,k)={d(d+1)k\ov 2 (k+d-1)}-{d(d-1)k\ov 2 (k+d-2)} \ .
\label{cvt2}
\ea
Then the difference
\ba
c_{\mathrm{C}H}(d,k + 2 d-4) - c_{\mathrm{C}S}(d-1,k) = 1 + {3(d-1)^2\ov k+d-3}\ ,
\label{pkj6}
\ea
corresponds to the central charge of the extra linear dilaton theory $\mathbb{R}_Q$.
Identifying this with $1+12 Q^2$ we read off the background charge as
\ba
Q_{d,k}= - {d-1\ov 2 \sqrt{k+d-3}}\ .
\label{bckk}
\ea

\no
Returning back to the discussion of the structure of the background metric and dilaton,
in the limit $b\to \infty$ and after shifting the level $k\to k+2 d-4$
we obtain
\ba
\mathrm{d}s^2 = 2(k+d-3) {\mathrm{d}b^2\ov 4b^2} + 2 (k+d-2)\,
\hat g_{ij}({\bf x};k)\,  \mathrm{d}x^i\, \mathrm{d}x^j\ ,
\label{hj3ex}
\ea
where
\ba
\hat g_{ij}({\bf x};k) = {k+d-3\ov k+d-2}\ g_{ij}(\infty,{\bf x};k+2 d-4)\ .
\label{kk4}
\ea
Similarly for the dilaton we have
\ba
 \mathrm{e}^{-2\Phi}= \mathrm{e}^{-2\Phi_0}
 \mathrm{e}^{-2 (d-1)x} {F({\bf x})\ov \sqrt{\hat \beta({\bf x};k)}}\ ,
\label{fo8}
\ea
with
\ba
\hat \beta({\bf x};k) = c_{d,k}\ \beta(\infty,{\bf x};k+2d -4)\ ,
\qq c_{d,k} = \left(k+d-3\ov k+d-2\right)^{d-1}\ ,
\label{jdw0}
\ea
where the value of the constant $c_{d,k}$ is dictated by the
right limiting behaviour as $b\to \infty$ as well as the relation \eqn{mkl2}.
We claim that the $(d-1)$-dimensional metric
\ba
\mathrm{d}s_{(d-1)}^2= 2 (k +d-2)\,  \hat g_{ij}({\bf x}) \, \mathrm{d}x^i\,  \mathrm{d}x^j\ ,\qq
\mathrm{e}^{-2\Phi_{(d-1)}}=
{F({\bf x})\ov \sqrt{\hat \b({\bf x};k)}}\ ,
\label{fdkexa}
\ea
correspond to the exact backgrounds for the compact coset model
$\mathrm{C}S_{d-1,k}$.

\subsection{General conclusions}

In conclusion, we schematically have the relation
\ba
\mathrm{C}H_{d,k+2 d-4} \quad {\rm at\ spatial\ infinity\ becomes}\quad
\mathrm{C}S_{d-1,k} \times \mathbb{R}_{Q_{d,k}}\ ,
\label{con1}
\ea
where we have emphasized the shift in the level $k$.
Besides the case of Euclidean-signature cosets we also note the relation
between the Minkowskian-signature cosets
\ba
\mathrm{CAdS}_{d,k+2 d-4} \quad {\rm at\ spatial\ infinity\ becomes}\quad
 \mathrm{CdS}_{d-1,k} \times \mathbb{R}_{Q_{d,k}}\ .
\label{con2}
\ea
The above two cases are obtained by considering the
spatial asymptotic regions of the higher-dimensional coset spaces and
the relation is between spaces of the same signature.
By considering a limiting procedure that involves the time variable
going to the infinite past (or future) we may obtain the following relation between
a Minkowskian-signature and a Euclidean-signature coset
\ba
\mathrm{CdS}_{d,k-2 d+4} \quad {\rm at\ temporal\ infinity\ becomes}\quad
 \mathrm{C}H_{d-1,k} \times \mathbb{R}_{\tilde Q_{d,k}}\ ,
\label{co3}
\ea
where in the latter case the Liouville field is timelike, so that
its central charge is $1-12 \tilde Q^2$, with
\ba
\tilde Q_{d,k}= - {d-1\ov 2 \sqrt{k-d+3}}\ .
\label{bckktime}
\ea

\boldmath
\section{The $d=3$ example} \label{d3}
\unboldmath

In this section we explicitly verify our claim \eqn{con1}
at the level of the semiclassical
and exact background fields for the three-dimensional case.

\boldmath
\subsection{The large-$k$ regime} \label{d3lk}
\unboldmath

For the three-dimensional case the metric and dilaton in the large-$k$
semiclassical limit were found for various patches of the global
space in \cite{BS1}. For our purposes it is convenient to use
global coordinates which have the appropriate ranges to cover the
entire manifold and according to their range they can describe all
different coset cases \eqn{ch}--\eqn{cs8} (for $d=3$). These variables
have a group-theoretical origin, relating to invariants of
the gauge subgroup and were constructed in \cite{BS3}. We have for
the metric
\be
\mathrm{d}s^2=2 k\left({\mathrm{d}\hat{b}^2\over
4(\hat{b}^2-1)}
+{\hat{b}-1\over \hat{b}+1}\ {\mathrm{d}\hat{u}^2\over 4\hat{u}(\hat{v}-\hat{u}-2)}-
{\hat{b}+1\over \hat{b}-1}\
{\mathrm{d}\hat{v}^2\over 4\hat{v}(\hat{v}-\hat{u}-2)}\right)\
\label{met3d}
\ee
and for the
dilaton
\be
\mathrm{e}^{-2\Phi} = \mathrm{e}^{-2\Phi_0} |(\hat{b}^2-1)(\hat{v}-\hat{u}-2)|\ .
\label{dil3d}
\ee
They are indeed of the forms
\eqn{smk1} and \eqn{fh4}, where, for later convenience,
we have used the notation $\hat b$ instead of just $b$.

\no
Depending on the ranges of the real variables
$\hat u,\hat v$ and $\hat b$, this background corresponds to either of the cosets
\eqn{ch}--\eqn{cs8} for $d=3$. Specifically \cite{BS3}
\ba \mathrm{CAdS}_3:
&&
\left\{|\hat{b}|>1, \hat{u}\hat{v}>0\right\}\quad {\rm or}
\quad \left\{|\hat{b}|<1, \hat{u}\hat{v}<0\
{\rm excluding}\ 0<\hat{v}<\hat{u}+2<2\right\}\ ,
\nonumber\\
\mathrm{CdS}_3 :  &&
\left\{|\hat{b}|>1, \hat{u}\hat{v}<0\
{\rm excluding}\ 0<\hat{v}<\hat{u}+2<2\right\}\
\quad {\rm or} \quad \left\{|\hat{b}|<1,\hat{u}\hat{v}>0\right\}\ ,
\nonumber\\
\mathrm{C}H_3 :  &&
\left\{|\hat{b}|>1,0<\hat{v}<\hat{u}+2<2\right\}\ ,
\label{rang3}\\
\mathrm{C}S_3 :  &&
\left\{|\hat{b}|<1,0<\hat{v}<\hat{u}+2<2\right\}\ .
\nonumber
\ea
The level $k$ appearing in \eqn{met3d}
is assumed to be
positive for the cosets $\mathrm{CAdS}_{d,k}$ and $\mathrm{C}H_{d,k}$. For the
$\mathrm{CdS}_{d,k}$ and $\mathrm{C}S_{d,k}$
we should flip its sign, i.e. $k\to -k$ in order to have the correct signature.
This is the origin of the negative sign in the levels appearing in \eqn{ch} and \eqn{ch8}.

\no
For later use note also that the above geometry has the following curvature invariants
(in all curvature invariants we compute in this paper we have not included the
overall $2k$ factor in the metric):
\begin{equation}\label{3dR}
  R=8\frac{3 + \hat{b}^2 + \hat{u}  - \hat{v} -
  b \left(\hat{u}+ \hat{v}\right)}{(\hat{b}^2-1)\left(\hat{v}- \hat{u} -2 \right)}
\end{equation}
and
\begin{equation}\label{3ddetratio}
  \frac{\det g_{\mu\nu}}{\det R_{\mu\nu}}=
  \frac{1}{64}(\hat{b}^2 - 1)\left(\hat{v}-  \hat{u} - 2\right)\ .
\end{equation}
The third invariant $R_{\mu\nu}\, R^{\mu\nu}=R^2/2$ is not independent
in this case.
Introduce finally the combination
\begin{equation}\label{3dK}
  K = 8 R \frac{\det g_{\mu\nu}}{\det R_{\mu\nu}}=3 + \hat{b}^2 + \hat{u}  - \hat{v} -
  \hat{b} \left(\hat{u}+ \hat{v}\right)^2\ ,
\end{equation}
which is also invariant, but not independent.
A {\it higher-order} in derivatives invariant, independent of the three canonical
curvature invariants of three-dimensional manifolds is\footnote{
In general a $d$-dimensional manifold has a number of scalars constructed out of the
metric and its derivatives. At most $d$ of such scalars are {\it functionally} independent.
The number of {\it algebraically} independent scalar invariants, not satisfying any polynomial
relation is much larger and was computed in \cite{haskins}, more than 100 years ago.
It can be proved, though it is intuitively obvious that there is an equivalent statement to say that
an invariant depends on the metric, the curvature tensor and its
covariant derivatives up to order $k$, called the order of the invariant.
Then in \cite{haskins} it was found that
the result for the number of zeroth order invariants, that is those built with the metric and
the curvature tensor is (see also for this case p. 145 of \cite{weinberg})
\ba
J_{d,0} = {1\ov 12} (d-2)(d-1)d (d+3)\ ,\qq d\geqslant 3\ .
\nonumber
\ea
For higher dimensional invariants that necessarily contain the $(k+2)$-th derivative of the metric or
equivalently the $k$-th convariant derivative of the curvature tensor, the result is
\ba
I_{d,k} = d\ {k+1\ov 2} {(d+k+1)!\ov (d-2)! (k+3)!}\ ,\qq k\geqslant 1 \ ,\quad d\geqslant 3\ .
\nonumber
\ea
There are in general differential relations among these invariants and the number of functionally
independent ones is less than $J_{d,0}$ and $I_{d,k}$ which should be thought of as upper bounds.
For $d=3$ we have $J_{3,0}=3$ and the invariants for our case are listed above.
We also have that $I_{3,2}=27$, so that the invariant $L$ in \eqn{3dL} should be related to
a linear combination of them.
}
\ba
\label{3dL}
  L=\frac{1}{3}\left(\frac{\Box}{4}-1\right)K
  =\hat{b}^2-1\ .
\ea
We restrict to the Euclidean-signature non-compact coset
$\mathrm{C}H_{3,k}$, we let the variable
$\hat{b}\to \infty$ and simultaneously trade it for $x$
as $\hat{b}= \mathrm{e}^{2 x}$, with $x\to \infty$, as in \eqn{bx}.
Then the variable $x$ in the metric \eqn{met3d} decouples whereas
in the dilaton \eqn{dil3d} it contributes the linear term $-2 x$.
Hence for generic values of $\hat u$ and $\hat v$ this is weak-coupling limit.
In order to interpret the remaining two-dimensional metric
let the
coordinate change
\ba
\hat{u}=-2 \sin^2\th \cos^2\phi\ ,\qq \hat{v}= 2 \sin^2\th \sin^2\phi\ .
\label{fcgh1}
\ea
This covers entirely the allowed range of the variables
$\hat{v}$ and $\hat{u}$ in \eqn{rang3}
and the two-dimensional part of the metric becomes
\ba
\mathrm{d}s^2_{(2)} = 2k (\mathrm{d}\th^2 + \tan^2\th \ \mathrm{d}\phi^2)\ .
\label{met2d}
\ea
In addition, the corresponding part of the dilaton is
\ba
\mathrm{e}^{-2\Phi_{(2)}}=\cos^2\th\ .
\label{dil2d}
\ea
The background \eqn{met2d} and $\eqn{dil2d}$ corresponds to the compact coset
$\mathrm{C}S_{2,k}$ (found for the coset $SL(2,\mathbb{R})/\mathbb{R}$ in \cite{wittenbh} from which
it is obtained by a simple analytic continuation and called the \emph{bell geometry}),
as advertized above.

\subsection{The exact background}

The exact, in the perturbative $1/k$ expansion,
expression for the metric has the form \cite{BSexa}
\be
\mathrm{d}s^2=2(k-2)\left(G_{\hat{b}\hat{b}} \, \mathrm{d}\hat{b}^2
+ G_{\hat{v}\hat{v}} \, \mathrm{d}
\hat{v}^2 +G_{\hat{u}\hat{u}} \, \mathrm{d}\hat{u}^2 +2G_{\hat{v}\hat{u}}\,
\mathrm{d}\hat{v}\, \mathrm{d}\hat{u}\right)
\label{ex3d}
\ee
with
\ba
G_{\hat{b}\hat{b}}& = & {1\over 4(\hat{b}^2-1)}\ ,
\nonumber\\
G_{\hat{u}\hat{u}} & = & {\b(\hat{b},\hat{u},\hat{v})\over 4\hat{u}(\hat{v}-\hat{u}-2)}
\left({\hat{b}-1\over \hat{b}+1} - {1\over k-1} {\hat{v}-2\over \hat{v}-\hat{u}-2}\right)\ ,
\nonumber\\
G_{\hat{v}\hat{v}}& = & -{\b(\hat{b},\hat{u},\hat{v})\over 4\hat{v}(\hat{v}-\hat{u}-2)}
\left({\hat{b}+1\over \hat{b}-1} + {1\over k-1} {\hat{u}+2\over \hat{v}-\hat{u}-2}\right)\ ,
\\
G_{\hat{v}\hat{u}} & = &
{1\over 4(k-1)} {\b(\hat{b},\hat{u},\hat{v})\over (\hat{v}-\hat{u}-2)^2}\ ,
\nonumber
\ea
where the function $\b(\hat{b},\hat{u},\hat{v})$
(compared to the general notation we have
adopted so far, we omit displaying $k$-dependence explicitly) is defined as
\ba
\b^{-1}(\hat{b},\hat{u},\hat{v})=1+{1\over k-1} {1\over \hat{v}-\hat{u}-2}
\left({\hat{b}-1\over \hat{b}+1} (\hat{u}+2)
-{\hat{b}+1 \over \hat{b}-1} (\hat{v}-2) -{2\over k-1}\right)\ .
\ea
In addition the exact dilaton is
\ba
 \mathrm{e}^{-2\Phi}= \mathrm{e}^{-2\Phi_0}
{|(\hat{b}^2-1)(\hat{v}-\hat{u}-2)|\ov \sqrt{\b(\hat{b},\hat{u},\hat{v})}}\ .
\label{dil33}
\ea
Indeed, \eqn{ex3d} and \eqn{dil33} are of the form \eqn{fl1} and \eqn{fh34}.
In the limit $x\to \infty$ ($\hat{b}=\mathrm{e}^{2 x}$),
the variable $x$ decouples as before and
similarly contributes a term $-2 x$ to the dilaton field. The remaining part of the metric,
after the shift $k\to k+2$ and the coordinate change \eqn{fcgh1}, becomes
\be
\mathrm{d}s^2_{(2)}= 2 (k+1)\left(\mathrm{d}\th^2 +(1-{1\ov k}
\tan^2\th)^{-1}\tan^2\th\ \mathrm{d}\phi^2\right)\ ,
\ee
whereas the corresponding exact dilaton becomes
\be
\mathrm{e}^{-2\Phi_{(2)}}= \left(1-{1\ov k} \tan^2\th\right)^{1/2} \cos^2\th\ .
\ee
This is nothing but the exact expressions for the metric and dilaton corresponding to
the $\mathrm{C}S_{2,k}$ coset \cite{BSexa,DVV}.

\boldmath
\section{The $d=4$ example}
\unboldmath

In this section we explicitly verify our claim \eqn{con1}
at the level of the semiclassical
and exact background fields for the four-dimensional case. This will be a highly
non-trivial check as we shall see.

\boldmath
\subsection{The large-$k$ regime}
\label{d4lk}
\unboldmath

For the four-dimensional cases the metric and dilaton in the large-$k$ semiclassical limit
were found for various patches of the global space in \cite{BS2}.
Again we use here the expression that can cover as before the global space for all
different cosets by appropriately restricting the ranges of the corresponding
variables \cite{BS3,BSexa}. The metric and the dilaton are
\ba
\mathrm{d}s^2 & = & 2k\Bigg(
{\mathrm{d}b^2\over 4(b^2-1)}+{b-1\over b+1} {\mathrm{d}u^2\over 4(v-u)(u-w)}
\nonumber\\
&& +\ {b+1\over b-1}{v-w\ov 4}\Big[{\mathrm{d}w^2\over (1-w^2)(u-w)}
-{\mathrm{d}v^2\over (v^2-1)(v-u)}\Big]\Bigg)\
\label{d4lkmet}
\ea
and
\ba
\mathrm{e}^{-2 \Phi}=\mathrm{e}^{-2\Phi_0} |(b^2-1)(b-1)(v-u)(w-u)|\ ,
\label{d4lkdil}
\ea
both of them of the form \eqn{met3d} and \eqn{dil3d}.
As in three dimensions, the signature is determined by the ranges of the four variables
$u,v,w$ and $b$. The analog of \eqn{rang3} is, as expected, more complicated and we restrict
our presentation on that for the coset $\mathrm{C}H_{4,k}$.
For the other cases the interested reader should consult \cite{BSexa}.
Since the metric is manifestly invariant under the interchange of the variables
$v$ and $w$ we may restrict to the range
\be
-1<v<u<w<1 \ ,\qq |b|>1 \ ,
\ee
with no loss of generality.
In the limit $b=\mathrm{e}^{2 x}\to \infty$ the coordinate $x$ decouples from the metric and it
contributes the linear term $-3 x$ to the dilaton. The remaining part of the metric has the form
\ba
\mathrm{d}s^2_{(3)} = 2 k\left[{\mathrm{d}u^2\over 4(v-u)(u-w)} + {v-w\ov 4}
\Bigg({\mathrm{d}w^2\over (1-w^2)(u-w)}
-{\mathrm{d}v^2\over (v^2-1)(v-u)}\Bigg)\right]\ .
\label{met3dd}
\ea
In addition the corresponding part of the dilaton is
\ba
\mathrm{e}^{-2 \Phi_{(3)}} = |(v-u)(w-u)|\ .
\label{dil3dd}
\ea
This limiting three-dimensional geometry has the following curvature invariants in one to one correspondence
with \eqn{3dR}--\eqn{3dL}:
\begin{equation}
\label{4dRlim}
  R=4\frac{1 - 2v w+ (u-v-w)^2}{(u - v) (u - w)}\ ,
\end{equation}
\begin{equation}
\label{4ddetratiolim}
  \frac{\det g_{\mu\nu}}{\det R_{\mu\nu}}=
  \frac{(u - v) (u - w)}{32} \ ,
\end{equation}
\begin{equation}\label{4dKlim}
  K = 8 R \frac{\det g_{\mu\nu}}{\det R_{\mu\nu}}=1 - 2v w+
  (u-v-w)^2
\end{equation}
and
\begin{equation}\label{4dLlim}
  L=\frac{1}{3}\left(\frac{\Box}{4}+1\right)K
  = - \left(u - v - w + 1\right)\left(u - v - w - 1\right)\ ,
\end{equation}
where we note a convenient flip of the relative sign between the two terms
in the higher-derivative invariant $L$ as compared with
\eqn{3dL}.
A relation of \eqn{met3d}, \eqn{dil3d} with \eqn{met3dd}, \eqn{dil3dd} is not apparent.
There is however a coordinate transformation that transforms the metric $\eqn{met3d}$ to
minus the metric \eqn{met3dd}.
This is constructed by comparing the corresponding
invariants \eqn{3dR}, \eqn{3ddetratio} and \eqn{3dL} with \eqn{4dRlim}, \eqn{4ddetratiolim}
and \eqn{4dLlim}. Since we want eventually the three-dimensional metrics
to be opposite to each other, we demand that the corresponding
$R$'s and $L$'s are opposite to each other, whereas the $K$'s are equal.
This provides
three algebraic relations between the two sets of coordinates from which we find
\begin{eqnarray}
\hat b&=& u - v - w \ ,
\nonumber\\
\hat{u}&=& -\frac{(1 + v) (1 + w)}{1 -u + v + w}\ ,
\label{cooch}\\
\hat {v}&=& \frac{(1 - v) (1 - w)}{1 +u - v - w}\nonumber \ .
\end{eqnarray}
The inverse transformation is
\begin{eqnarray}
u&=&\hat{b}+
\frac{\hat{b}}{2}\left(\hat{u} - \hat{v}\right)-\frac{1}{2}\left(\hat{u}
+ \hat{v}\right)\ ,
\nonumber\\
v + w &=&  \frac{\hat{b}}{2}\left(\hat{u} - \hat{v}\right) -\frac{1}{2}\left(\hat{u}
+ \hat{v}\right)\ ,
\label{coochinv}\\
v u&=& \frac{\hat{b}}{2}\left(\hat{u} +
\hat{v}\right)-\frac{1}{2}\left(\hat{u} - \hat{v}\right)-1  \ .
\nonumber
\end{eqnarray}
Then the metric (\ref{met3d}) indeed
transforms into minus \eqn{met3dd} and accordingly the dilaton.
It is worth stressing that the transformation above would have been hard to find
without the method of comparing invariants.

\subsection{The exact background} \label{d4eb}

We will now verify to all orders in $1/k$ that the asymptotic
region of the four-dimensional coset under consideration indeed
matches the three-dimensional coset plus a free boson. The exact
metric and dilaton are \cite{BSexa}
\ba
\mathrm{d}s^2 & =  & 2(k-3)
\Big(G_{bb}\, \mathrm{d}b^2 +G_{uu} \, \mathrm{d}u^2 +G_{vv}\,  \mathrm{d}v^2 +G_{ww}\,  \mathrm{d}w^2
\nonumber\\
&& +\ 2 G_{uv} \, \mathrm{d}u \, \mathrm{d}v +2 G_{uw}\,
\mathrm{d}u\,  \mathrm{d}w +2 G_{vw} \, \mathrm{d}v \, \mathrm{d}w\Big)\ ,
\label{ex4d}
\ea
where
\ba
G_{bb}& = & {1\over 4(b^2-1)}\ ,
\nonumber\\
G_{uu} & = & {\b(b,u,v,w)\over 4(u-w)(v-u)}\left[{b-1\over b+1}-{1\over k-2}
{(v-w)^2\over {(v-u)(u-w)}} \left(1-{1\over k-2} {b+1\over b-1}\right)\right]\ ,
\nonumber\\
G_{vv}& = & -{(v-w)\ \b(b,u,v,w)
\over 4(v^2-1)(v-u)}\Bigg[{b+1\over b-1}-{1\over k-2}
{1\over (v-u)(u-w)}
\nonumber\\
&&\hskip 0 cm \times \left(1-u^2
 +\left({b+1\over b-1}\right)^2 (v-u)(v-w)  +{1\over k-2}
{b+1\over b-1} {(1+v^2)(u+w)-2v(1+ u w)\over v-w}\right)\Bigg]\ ,
\nonumber\\
G_{ww} & = & {(v-w)\ \b(b,u,v,w)\over 4(1-w^2)(u-w)}
 \Bigg[{b+1\over b-1}-{1\over k-2}
{1\over (v-u)(u-w)}
\nonumber\\
&&\hskip 0 cm \times \left(1-u^2
  +\left({b+1\over b-1}\right)^2 (u-w)(v-w)
    -{1\over k-2}
{b+1\over b-1} {(1+w^2)(u+v)-2w(1+uv)\over v-w}\right)\Bigg]\ ,
\nonumber\\
G_{uv} & = & {\b(b,u,v,w)\over 4(k-2)(v-u)^2}\left(1-{1\over k-2} {b+1\over b-1}
{v-w\over u-w}\right)\ ,
\\
G_{uw} & = & {\b(b,u,v,w)\over 4(k-2)(u-w)^2}\left(1-{1\over k-2} {b+1\over b-1}
{v-w\over v-u}\right)\ ,
\nonumber\\
G_{vw} & = & {1\over (k-2)^2} {b+1\over b-1} {\b(b,u,v,w)\over
4(v-u)(u-w)}\ ,
\nonumber
\ea
with the function $\b(b,u,v,w)$ (again, we omit displaying $k$-dependence
explicitly) being defined as
\ba
&&\b^{-1}(b,u,v,w)  =  1+{1\over k-2} {(v-w)^2
\over (v-u)(w-u)} \Bigg[{b+1\over b-1}+{b-1\over b+1} {1-u^2\over (v-w)^2}+
\nonumber\\
&&
{1\over k-2}\left({vw+u(v+w)-3 \over (v-w)^2}
- \left({b+1\over b-1}\right)^2\right)\Biggr]
+{2\over (k-2)^3} {b+1\over b-1} {vw-1 \over (v-u)(u-w)} .
\ea
The dilaton field is
\be
\mathrm{e}^{-2 \Phi}=\mathrm{e}^{-2\Phi_0}
{|(b^2-1)(b-1)(v-u)(w-u)|\ov \sqrt{\b(b,u,v,w)}}\ .
\ee
This background in the limit $b\to \infty$ gives rise to a three-dimensional metric,
which after the shift $k\to k+4$, can be written in the form
\ba
 \mathrm{d}s^2_{(3)}\!\! & = &\!\!  2 (k+1) \beta(\infty,u,v,w)
 \Bigg[\left(A_1+{k+1\ov (k+2)^2} A_2\right) \mathrm{d}u^2
+ \left(B_1+{B_2\ov k+2} + {B_3\ov (k+2)^2} \right) \mathrm{d}v^2
\nonumber\\
&& + \left(C_1+{C_2\ov k+2} + {C_3\ov (k+2)^2} \right) \mathrm{d}w^2
 + \ 2\left({D_2\ov k+2} + {D_3\ov (k+2)^2} \right) \mathrm{d}u\, \mathrm{d}v
\label{metroo}\\
&& +\ 2\left({E_2\ov k+2} + {E_3\ov (k+2)^2} \right) \mathrm{d}u\, \mathrm{d}w
 +  2 {Z_3\ov (k+2)^2}  \mathrm{d}v\, \mathrm{d}w \Bigg]\ ,
\nonumber
\ea
where the various entries are
\ba
&& A_1 = {1\ov 4(u - w)(v - u)}\ ,\qq A_2 = - {(v - w)^2\ov 4(v - u)^2(u - w)^2}\ ,
\nonumber\\
&& B_1 = -{v - w\ov 4(v^2 - 1)(v - u)} \ ,\qq
B_2 = {(v - w)\left[1 - u^2 + (v - u)(v - w)\right]\ov 4(v^2 - 1)(v - u)^2(u -w)} \ ,
\nonumber\\
&&
B_3 = {(1 + v^2)(u + w) - 2 v (1 + u w)\ov 4(v^2 - 1)(v - u)^2 (u - w)}
\ ,
\nonumber\\
&& C_1 = -{w - v\ov 4(w^2 - 1)(w - u)} \ ,\qq
C_2 = {(w - v)\left[1 - u^2 + (w - u)(w - v)\right]\ov 4(w^2 - 1)(w - u)^2(u -v)} \ ,
\nonumber\\
&&
C_3 = {(1 + w^2)(u + v) - 2 w (1 + u v)\ov 4(w^2 - 1)(w - u)^2 (u - v)} \ ,
\\
&& D_2 = {1\ov 4(v - u)^2}\ , \qq D_3 = -{v-w\ov 4(v - u)^2 (u - w)}\ ,
\nonumber\\
&& E_2 = {1\ov 4(w - u)^2}\ ,\qq E_3 = -{w-v\ov (w - u)^2 (u - v)}\ ,
\nonumber\\
&& Z_3 = {1\ov 4(v - u)(u - w)}\ .
\nonumber
\ea
With these definitions the metric \eqn{metroo} can be recast as
\ba
\mathrm{d}s^2_{(3)} & = & 2 (k+2) \beta_\infty(u,v,w)\Bigg[\left(A_1+{2 A_1+A_2\ov k+1}+ {A_1\ov (k+1)^2}
\right) \mathrm{d}u^2
\nonumber\\
&& + \left(B_1+{2 B_1+B_2\ov k+1} + {B_1+B_2+B_3\ov (k+1)^2} \right) \mathrm{d}v^2
\nonumber\\
&& + \left(C_1+{2 C_1+C_2\ov k+1} + {C_1+C_2+C_3\ov (k+1)^2} \right) \mathrm{d}w^2
 + \ 2\left({D_2\ov k+1} + {D_2+D_3\ov (k+1)^2} \right) \mathrm{d}u\, \mathrm{d}v
\nonumber\\
&& +\ 2\left({E_2\ov k+1} + {E_2+E_3\ov (k+1)^2} \right) \mathrm{d}u\, \mathrm{d}w
 +  2 {Z_3\ov (k+1)^2} \ \mathrm{d}v\, \mathrm{d}w \Bigg]\ ,
\label{metrer}
\ea
where $\b_\infty$ is defined as
\be
\b^{-1}(\infty,u,v,w)= \left(k+1\ov k+2\right)^3\ \b^{-1}_\infty(u,v,w)\ ,
\ee
where the overall constant is precisely $c_{4,k}$ in \eqn{jdw0}.
We can now check that the above expression for the metric can be obtained from
the metric \eqn{ex3d} under the transformation \eqn{cooch} and the sign change $k\to -k$
(note that precisely $\b_\infty(u,v,w) = \b(\hat b,\hat u,\hat v)$).
It is remarkable that the coordinate transformation \eqn{cooch} does not receive
$1/k$-corrections, a fact that is attributed to the group-theoretical nature of the
coordinates we are using.

\section{Exactly marginal perturbations and parafermions} \label{d3eb}

So far we have explicitly demonstrated that in three
and four dimensions, there exists a limiting procedure
in the $\mathrm{C}H_d$ coset
involving a radial coordinate taken to spatial infinity,
in which this coordinate decouples
and the remaining lower-dimensional background corresponds to the coset $\mathrm{C}S_{d-1}$.

\no
We can recast this observation from a slightly different perspective by saying that
the conformal sigma models $\mathrm{C}H_d$ and  $\mathrm{C}S_{d-1}\times \mathbb{R}_Q$
have target spaces and dilaton backgrounds that coincide in some asymptotic corner. Hence,
these two sigma models are both solutions of the full beta-function equations with
common asymptotics. One therefore expects to find a continuous family of exact solutions
$\mathrm{C}\Sigma_d(\lambda)$ interpolating between $\mathrm{C}S_{d-1}\times \mathbb{R}_Q$
(at $\lambda = 0$) and $\mathrm{C}H_d$ (at $\lambda = 1$), and sharing the above
asymptotics. The larger the parameter $\lambda$, the wider the region of
$\mathrm{C}\Sigma_d(\lambda)$ coinciding with $\mathrm{C}H_d$ and for extreme deformation,
one recovers the complete $\mathrm{C}H_d$. For infinitesimal deformation $\delta\lambda$,
$\mathrm{C}\Sigma_d(\delta\lambda)$ provides the leading correction to
$\mathrm{C}S_{d-1}\times \mathbb{R}_Q$ that must coincide with the leading correction to the
asymptotics of $\mathrm{C}H_d$.

\no
Here we will compute this leading correction
and show that from the  $\mathrm{C}S_{d-1}\times \mathbb{R}_Q$
viewpoint, it corresponds to a marginal perturbation involving the non-Abelian parafermions
of the $\mathrm{C}S_{d-1}$ coset, for $d\geqslant 4$
or the Abelian ones of the $\mathrm{C}S_2$, for $d=3$, dressed with the Liouville field
of $\mathbb{R}_Q$.

\subsection{General considerations}

We consider again the expansion of the exact metric \eqn{fl1} of $\mathrm{C}H_d$, but now we keep the
leading correction in the $b\to \infty$ limit. We get an expression of the form
\ba
\mathrm{d}s^2 = 2(k+d-3) {\mathrm{d}b^2\ov 4b^2} + 2 (k+d-2) \hat g_{ij}({\bf x};k)\,
\mathrm{d}x^i\, \mathrm{d}x^j + {1\ov b}
\xi^i S_{ij} \xi^i + {\cal O}\left(1\ov b^2\right)\ ,
\label{hj3ex1}
\ea
The decoupled variable $b$ corresponds to a free boson. Instead of the change of
variable \eqn{bx}, in the discussion of this section it helps to absorb
the $k$-dependence into the transformation as
\be
b =  \mathrm{e}^{2 x} = \mathrm{e}^{-a_{d,k} \Phi} \ ,\qq a_{d,k}= -{1\ov \sqrt{k+d-3}}\ ,
\label{ber}
\ee
so that the boson $\Phi$ is properly normalized
and has a background charge given by \eqn{bckk}.

\no
Now we turn our attention to the leading correction term and show that
it this is indeed an exactly
marginal perturbation of the $\mathrm{C}S_{d-1}\times \mathbb{R}_Q$.
This correction term contains, via the factor $1/b$ a vertex operator of the form
\be
V_{d,k} = \mathrm{e}^{a_{d,k}\Phi}\ ,
\ee
which has conformal dimension
\ba
\D_{d,k}=\bar \D_{d,k} = -{a_{d,k}^2\ov 2} + Q_{d,k} a_{d,k} =
{d-2\ov 2 (k+d-3)}\ ,
\label{conff}
\ea
where in the second equality we have substituted the values for $a_{d,k}$
and $Q_{d,k}$ using \eqn{ber} and \eqn{bckk}.
The bilinear $\xi^i S_{ij} \xi^j$, where $\xi^i$ are differentials and $S_{ij}$ a
non-constant, in general,
matrix multiplying this vertex, corresponds
to the classical parafermions bilinear of the coset
$\mathrm{C}S_{d-1,k}$ theory, as we next explain, after a
short introduction on those aspects of parafermions, both classical
and quantum, that are necessary in this paper.

\no
Parafermionic quantum algebras were first introduced and analyzed in \cite{qmparaf}
and it turns out that they correspond to natural chiral and antichiral objects
in the $SU(2)/U(1)$ conformal coset theory.\footnote{
The analogous construction for the coset $SL(2,\mathbb{R})/\mathbb{R}$ led
to the non-compact parafermions \cite{lykken}.}
Their classical counterparts, called classical parafermions,
of a general coset $G/H$ theory were introduced later in
\cite{BCR,parafNON} as gauge-invariant objects of the theory.
They essentially arise
from the currents of the WZW theory for the group $G$ with coset indices
dressed with Wilson lines involving the gauge fields that
eventually are integrated out in the path integral in the gauged WZW action.
Denoting by Latin (Greek) letters subgroup $H$ (coset $G/H$) indices,
we have a set of holomorphic parafermions $\Psi^\a(\s^+)$
and a set of antiholomorphic ones
$\bar \Psi^\a(\s^-)$.
The
non-trivial braiding properties are reflected in the classical Poisson bracket
algebra they obey \cite{BCR}
\ba
\{ \Psi_\a(x),\Psi_\b(y)\} & = & - {k\ov \pi} \d_{\a\b}
\d^\prime (x-y)
- f_{\a\b\g} \Psi_\g (y) \d(x-y)
\nonumber \\
& & - {\pi\ov 2k}
f_{c\a\g} f_{c\b\d}~ \e(x-y) \Psi_\g(x) \Psi_\d (y) ~ ,
\label{poipar}
\ea
where the antisymmetric step function
$\e(x-y)$ equals $+1~ (-1)$ if $x>y$ ($x<y$).
The last term in \eqn{poipar} is responsible
for their non--trivial
monodromy properties and unusual statistics.
In the above, $x$ and $y$ denote world-sheet light-cone
variables such as $\s^+$ and $\s'^+$, so that
the Poisson brackets in \eqn{poipar} are evaluated at
equal light-cone time $\s^-$.
In addition, conformal transformations are generated by
$T_{++}=-{\pi\ov 2 k} \Psi_\a\Psi_\a$.
A similar algebra to \eqn{poipar} and statements
are valid for the antiholomorphic parafermions $\bar \Psi^\a$ as well.

\no
The classical parafermions have dimension one, but quantum mechanically
they receive $1/k$-corrections.
In \cite{bardakciNONabel}
the above Poisson algebra was promoted at the level of an
OPE conformal algebra and various
consistency conditions were checked extensively. In particular, it
was found that the anomalous dimension of the parafermions
is completely dictated by the structure of the
braiding last term in the Poisson algebra \eqn{poipar}.
For the general case the dimension matrix is
\ba
\D_{\a\b}=\bar \D_{\a\b}= \d_{\a\b}-{f_{c\g\a} f_{c\g\b}\ov k+g_H}\ ,
\label{dimm}
\ea
where the dual Coxeter number $g_H$ for the subgroup is $f_{acd}f_{bcd} = g_H \d_{ab}$.
There is an alternative way of seeing that this is the conformal dimension of the
parafermions. Recall, that they are essentially the generators $J_\a$ of the currents with
coset indices. Then with respect to the energy--momentum tensor of the coset theory
\ba
T_{G/H} =T_G-T_H= {:J^AJ^A:\ov k+g_G}- {:J^aJ^a:\ov k+g_H}\ ,
\ea
where we normal order the current bilinears  according to the prescription in \cite{bais},
we compute that
\ba
J_\a(z) T_{G/H}(w) = {\D_{\a\b}J_\b(w)\ov (z-w)^2} + {\rm regular}\ ,
\ea
where the dimension matrix $\D_{\a\b}$ is precisely that in \eqn{dimm}.
Hence, we may think that the parafermions $\Psi_\a$ essentially inherit the
dimension of the coset generators $J_\a$ with the respect to the
energy--momentum tensor $T_{G/H}$.

\no
In our case the group is $G=SO(d)$ and the subgroup $H=SO(d-1)$.
According to our normalizations and using the representation matrices for $SO(d)$,
\be
(t_{AB})_{CD}= \sqrt{2} \d_{C[A} \d_{B]D} \ ,
\ee
we compute the structure constants as
\be
SO(d):\phantom{xxxx}
f_{AB,CD,KL}={1\ov \sqrt{2}} \d_{BC}\d_{AK}\d_{DL} + {\rm antisymmetric}\ ,
\ee
from which the dual Coxeter number is $g_{SO(d)}=d-2$.
We split the indices as $A=(0,i)$, with $i=1,2,\dots , d-1$, being subgroup $SO(d-1)$
indices.
Then the algebra of the parafermions \eqn{poipar} becomes
\be
\{ \Psi^i(x),\Psi^j(y)\} =  -{k\ov  \pi} \d_{ij} \d^\prime (x-y)
-{\pi\ov 4 k} \e(x-y) \Big[ \d_{ij} \Psi(x)\cdot \Psi(y)
- \Psi_j(x) \Psi_i(y) \Big] ~ .
\label{poi}
\ee
The absence of linear terms in $\Psi^i$
on the right-hand side
is due to the simple fact that $SO(d)/SO(d-1)$
is a symmetric space.
Thus, structure constants involving only coset space indices are zero.
In addition, we have that the bilinear in the structure constants that appears in
the dimension formula \eqn{dimm} is
$\ha f_{mn,0k,0i} f_{mn,0k,0j}=\ha (d-2)\ \d_{ij}$, where the factor of $\ha$
is to avoid overcounting.
Hence, the dimension metric for the case of the $\mathrm{C}S_{d-1,k}$ coset theory
is diagonal, i.e. $\D_{ij}= \D^{\mathrm{C}S_{d-1}}_{k} \d_{ij}$
with
\be
\D_k^{{\mathrm{C}S_{d-1}}}= \bar \D_k^{{\mathrm{C}S_{d-1}}} = 1-{d-2\ov 2(k+d-3)}\ .
\ee
Taking this into account we see that the anomalous dimension of the
parafermions exactly cancels that of the vertex oparator \eqn{conff}.
Hence the perturbation is indeed marginal.
Note that for $d=3$, corresponding to the $\mathrm{C}S_2$ model,
we get the correct dimension of the quantum parafermions of \cite{qmparaf}.

\no
Finally, we note that
the classical parafermions of the $\mathrm{C}S_{d-1}$ coset theory have the form
\be
\Psi^i  =
{ik\ov \pi}\ \psi^j \ h_+^{ji} \ ,\qq \bar\Psi^i  =
{ik\ov \pi} \ \bar\psi^j \ h_-^{ji} \ ,
\label{equff}
\ee
with
\be
h_\pm^{-1} = {\rm P}  \mathrm{e}^{- \int^{\s^\pm} A_\pm}\ ,
\label{hphm}
\ee
where ${\rm P}$ stands for path ordering and note that $h_\pm\inv = h_\pm^T$.
The $\psi^i$ and $\bar\psi^j$'s are local expressions of the space variables and
first order in the world-sheet derivatives $\del_+$ and $\del_-$, respectively
and precisely those that appear in the correction term.
The path ordered exponentials give to them the non-local character in the space variables.
With these definitions the $\Psi^i$ and $\bar \Psi^i$ are chiral
and antichiral, respectively. The $A_\pm$'s are nothing
but the $SO(d-2)$ gauge fields, taking their
on-shell values assumed when they are integrated out in the gauged WZW action.
Their specific expressions are not relevant to this paper.

\no
With the above general considerations we claim that the perturbation to
the action corresponding to the metric \eqn{hj3ex1} takes the form
\ba
\d \cL \sim k\ V_{d,k} \psi^i \bar \psi^i = k\ V_{d,k} \Psi^i \bar \Psi^i\ ,
\label{prth}
\ea
where the non-trivial Wilson factors drop out since, on-shell, $h_+=h_-$.
In the point particle limit the $\psi^i$ and $\bar \psi^i$'s become the differentials
$\xi^i$ and $S_{ij} \xi^j$, respectively, thus reproducing the metric perturbation
in \eqn{hj3ex1}.
This kind of marginal perturbation, involving dressed parafermions, appeared
first, to the best of our knowledge, in \cite{PS05}. In there, the
background corresponding to NS5-branes distributed on an ellipsis was
explicitly constructed. The small deformation around the uniform distribution
of the NS5-branes on a circle, described by an orbifold of the
$SL(2,\mathbb{R})/\mathbb{R} \times SU(2)/U(1)$ direct product
conformal cosets \cite{sfet1}, is precisely an exactly marginal deformation
involving the, appropriately dressed, parafermions of \cite{qmparaf}.

\no
In the rest of this section we verify the above general statements for the three-
and four-dimensional cases in which explicit results are available. Before proceeding we should
stress that the above considerations prove rigourously (i.e. at any finite value of $k$)
and generally (i.e. for any $d$) that $\delta \mathcal{L}$ in \eqn{prth} is a marginal
operator of $\mathrm{C}S_{d-1}\times \mathbb{R}_Q$. However (\romannumeral1)
neither these considerations prove that $\delta\mathrm{d}s^2$ in \eqn{hj3ex1} always
originate from \eqn{prth} -- this is the purpose of next paragraphs for $d=3,4$;
(\romannumeral2) nor do they demonstrate that the marginal operator \eqn{prth}
is integrable (i.e. exact). This property is here inferred from the fact that
this operator connects two exact CFT's, namely
$\mathrm{C}S_{d-1}\times \mathbb{R}_Q$ and $\mathrm{C}H_{d}$. A proof based on genuine
CFT techniques is beyond the scope of the present paper.

\boldmath
\subsection{The $d=3$ example and Abelian parafermions}
\unboldmath

Let's consider the three-dimensional metric \eqn{met3d} in the large-$b$ limit,
keeping however the first correction in the $1/b$ expansion. We find
that
\be
\mathrm{d}s^2= 2k \mathrm{d}x^2 + \mathrm{d}s^2_{(2)}
- 4k \mathrm{e}^{-2 x} \left[ \cos2\phi \left(\mathrm{d}\th^2-\tan^2 \th\, \mathrm{d}\phi^2\right)
-4 \tan\th\, \sin\phi\, \cos\phi\, \mathrm{d}\th \, \mathrm{d}\phi\right]\ .
\label{cor2d}
\ee
This should be an exactly marginal perturbation as the full three-dimensional model
is conformal
and therefore it should
be such at every order in perturbation theory in powers of $1/b$.
so that it manifestly corresponds to a marginal perturbation.
Our aim is to rewrite the $\s$-model action corresponding to
metric \eqn{cor2d} in terms of natural objects in the $\mathrm{C}S_{2,k} \times \mathbb{R}_{Q_{3,k}}$
CFT.
For the $\mathrm{C}S_{2,k}$ factor the natural objects are the classical parafermions
\cite{BCR}. The semiclassical expressions for the chiral
parafermions in terms of space variables are (we ignore a factor involving $k$)
 \be
 \Psi_\pm =
 (\del_+\th \mp i \tan\th \del_+\phi)
\mathrm{e}^{\mp i (\phi +\phi_1)}\ ,
 \label{ncpar}
 \ee
where the phase is
\be
\phi_1  =
-\ha \int^{\s^+}\! J^1_+\mathrm{d}\s^+ + \ha \int^{\s^-} J_-^1 \mathrm{d}\s^-\ ,\qq
J^1_\pm = \tan^2\th \del_\pm \phi\ .
\label{phi}
\ee
The phase obeys on-shell the condition $ \del_+\del_- \phi_1 = \del_-\del_+ \phi_1$
and is well defined, due to the classical equations of motion.
Similarly, the expressions for the antichiral ones are
 \be
 \bar \Psi_\pm =
 (\del_-\th \pm i \tan\th \del_-\phi)\
\mathrm{e}^{\pm i (\phi -\phi_1)}\ .
\label{ncbpar}
\ee
It is easy to show that the correction third term in \eqn{cor2d}
can be reproduced by adding to the sigma-model action based on the
unperturbed background $\mathrm{C}S_{2,k} \times \mathbb{R}_{Q_{3,k}}$ the term
 \be
 \d \cL = -2 k\ V_{3,k} \left(\Psi_+ \bar \Psi_- + \Psi_- \bar \Psi_+\right)\ ,
\label{delScir}
 \ee
as a perturbation. This takes the form \eqn{prth} and
is a $(1,1)$ marginal perturbation as we have explicitly
shown in the general case.

\boldmath
\subsection{The $d=4$ example and non-Abelian parafermions}
\unboldmath

It turns out to be more convenient to trade the coordinates $(\hat{b},\hat{u},\hat{v})$
introduced in Sec.~\ref{d3lk} for the angular coordinates $(\theta,\phi,\omega)$ defined as
\be
\hat{b}=\cos 2\th\ ,\qq \hat{u} = -2 \sin^2\phi \sin^2\om \ , \qq \hat{v} = 2 \cos^2\phi\ .
\label{angtr}
\ee
This transformation completely covers the range of variables corresponding to the
$\mathrm{C}S_3$ compact coset in \eqn{rang3}.
The background metric is 
\ba
\mathrm{d}s^2_{(3)} = \mathrm{d}\th^2 +
\tan^2\th (\mathrm{d}\om + \tan\om \, \cot \phi\,  \mathrm{d}\phi)^2 + {\cot^2\th \ov
\cos^2\om} \mathrm{d}\phi^2 ~ ,
\label{ds3}
\ea
whereas the dilaton reads
\begin{equation}
\label{3ddiangl}
  \mathrm{e}^{-2\Phi_{(3)}} =  \mathrm{e}^{-2\Phi_0}\,
  \sin^2 2\theta \, \sin^2 \phi\,  \cos^2\omega  \ .
\end{equation}

%
\no
The chiral parafermions were explicitly computed in \cite{BakSfe}, a
work on universal aspects of string theories, precisely in terms of
the variables in the metric \eqn{ds3}. To conveniently present them
for our purposes, let's introduce the forms
\begin{eqnarray}
\xi^1 & = & {\cot\th\ov \cos\om}\,
\mathrm{d}\phi\ ,
\nonumber\\
\xi^2
 & = &
\cos \omega \,\mathrm{d}\theta - \tan \theta \, \cot \phi \,
\frac{\sin^2 \omega}{\cos \omega}\, \mathrm{d}\phi - \tan \theta
\,\sin \omega\,\mathrm{d} \omega\ ,
\label{foor}\\
 \xi^3
 & = &
\sin \omega \,\mathrm{d}\theta + \tan \theta \, \cot \phi \, \sin
\omega\, \mathrm{d}\phi + \tan \theta \,\cos \omega\,\mathrm{d}
\omega
\nonumber
\end{eqnarray}
and define the $\xi^i_\pm $'s via the expansion
\ba
\xi^i = \xi_+^i \mathrm{d}\s^+ + \xi_-^i \mathrm{d}\s^-\ ,\qq i=1,2,3\ .
\label{jf4}
\ea
Then in this basis, the chiral parafermions take the form
\ba
\psi^i = \xi^i_+ \ ,\qq i=1,2,3\ .
\label{hl2}
\ea
As a check, one may
verify that $\Psi^i\Psi^i = \psi^i \psi^i =\xi^i_+\xi^i_+$
is indeed proportional to the component $T_{++}$ of the energy--momentum
tensor corresponding to a sigma-model with metric
\eqn{ds3}. For antichiral ones, we have the slightly different form
\ba
\label{DDDbarX}
\bar \psi^i = S_{ij} \xi_-^i \ ,
\ea
where $S$ is an orthogonal matrix, not connected to the identity, given by
\ba
S=\begin{pmatrix}
 -\cos 2\phi &  \sin 2 \phi & 0
 \\
\sin 2 \phi&  \cos 2 \phi &  0
 \\
0  & 0  & 1
\end{pmatrix} \ .
\ea
Again we note that $\bar \Psi^i\bar \Psi^i =\bar \psi^i \bar\psi^i =\xi^i_-\xi^i_-$
is indeed proportional to the component $T_{--}$ of the energy--momentum
tensor corresponding to a sigma-model with metric
\eqn{ds3}.

\no
Let us now focus on the $\mathrm{C}H_4$ coset. In the asymptotic region where $b= \mathrm{e}^{ 2x}$
is large,
the background fields read (see Eqs. (\ref{d4lkmet}) and (\ref{d4lkdil})):
\be
\label{4dmetnlo}
\mathrm{d}s^2  =2k {\rm d}x^2 + {\mathrm{d}s^2_{(3)}}+ {\d \mathrm{d}s^2} \ ,
\ee
where
\ba
\label{4deldmetnlo}
\delta{\rm d}s^2 = k \mathrm{e}^{-2x}
\left[\frac{{\rm
d}u^2}{4(u-v)(u-w)}+\left(\frac{(w-v){\rm
d}v^2}{4\left(1-v^2\right)(u-v)}+\frac{(v-w){\rm
d}w^2}{4\left(1-w^2\right)(u-w)}\right)\right] \ .
\ea
With the help of the coordinate transformation (\ref{coochinv})
together with (\ref{angtr}), we can recast the subleading term
\eqn{4deldmetnlo} as
\ba
\label{3dmetdef}
\delta \cL =4 k\ V_{4,k}\psi^i \bar \psi^i= 4 k\ V_{4,k}\Psi^i \bar \Psi^i\ ,
\ea
again of the form \eqn{prth}.
This is a $(1,1)$ marginal perturbation as we have explicitly
shown in the general case.

\section{Discussion}

\no
The main physical outcome of this work is the appearance of exact
$d$-dimensional backgrounds $\mathcal{B}$, target spaces of gauged
WZW models, whose $(d-1)$-dimensional ``boundaries''
$\partial\mathcal{B}$ are also exact CFT's. Supplemented
with an extra free field with background charge, the theory on $\partial\mathcal{B}$
admits a truly marginal deformation that allows to reconstruct the theory
on $\mathcal{B}$. In a CFT language one can say that the background $\mathcal{B}$
is build as the dynamical promotion of the marginal
deformation line to an extra dimension.

\no
Similar situations have been analyzed in the literature both
for marginal and for relevant deformations, aiming at promoting the
spectral parameter or the scale of a RG flow to a genuine space coordinate. What is remarkable about
the above results is that,
rephrased in more physical terms,
they mean that the theory on $\mathcal{B}$ is
generated from its own asymptotic data.
It is very interesting to relate the representation theory of the group $SO(d,1)$,
appropriate to the non-compact coset $C\!H_d$, to that of the group $SO(d)$, appropriate for the
compact coset $C\!S_{d-1}$. We expect that representations similar to the principal series
representation of $SL(2,\mathbb{R})$ will have a limit such that they reduce to $SO(d)$
representations appropriate for the $C\!S_{d-1}$ compact coset.
In that spirit, an interesting issue worth the investigation is
the propagation of fields in the
background $\mathcal{B}$ in relation to asymptotic or initial data in the
remote region $\partial\mathcal{B}$. This might help in reconsidering in a
truly string framework ideas that have been sofar explored only in field theory.

\no
The backgrounds at hand, $\mathcal{B}$, $\partial\mathcal{B}$, \dots form hierarchies of
gauged WZW models on orthogonal groups.
The corresponding target spaces can be Euclidean or Minkowskian,
where the $\partial\mathcal{B}$ is either time-like
or space-like. A space-like $\partial\mathcal{B}$ is interpreted as a collection of
data in remote time that evolve toward the future. It would be interesting to analyze the
potential cosmological applications of the backgrounds at hand, which provide a CFT
generalization of the FRW solutions: in the FRW universes, any spatial section is a
maximally symmetric solution
of Einstein's equations, whereas in our case only initial data provide a good CFT.

\no
As a bonus, our present analysis enables us to answer a long-standing question, at least for
orthogonal groups: how to obtain
the gauged WZW model $G/H$ as the endpoint of a marginal deformation, much like $G/U(1)$
appears as an extreme deformation of $G$ under the appropriate current--current operator.
It will be interesting to investigate similar issues in conformal coset theories
based on other non-Abelian groups.
In that respect, we note the explicit results in \cite{Lugo} for the $SU(2,1)/U(2)$
and $SU(2,1)/SU(2)$ conformal coset theories.

\no
Finally, we would like to stress once more the emergence
of parafermions as building blocks of exactly marginal operators, when
appropriately dressed. Our proof that these operators are indeed integrable is indirect
and relies on the independent observation that they generate a line of
continuous deformation with vanishing beta-functions to all orders. A proof based on
genuine CFT techniques would require mastering of (non-)Abelian quantum parafermions,
which is a notoriously difficult subject. Notice also that marginal operators were usually
thought of as products of holomorphic times antiholomorphic currents, absent in gauged WZW
(by lack of any residual symmetry).
To our knowledge, parafermion-based marginal operators
appeared only recently in \cite{PS05},
where their effect was to deform the circular shape of the NS5-brane
distribution that generates the background.

\vskip .4 in
\centerline{ \bf Acknowledgments}

\no
The author P.M. Petropoulos thanks I. Bakas for scientific discussions.
He also acknowledges the University of Patras for
kind hospitality and financial support as well as financial support by
the INTAS contract 03-51-6346, by the EU under the
contracts MRTN-CT-2004-005104, MRTN-CT-2004-503369 and MEXT-CT-2003-509661
and by the Swiss National Science Foundation.\\
The author K. Sfetsos
acknowledges partial support
provided through the European Community's
program ``Constituents, Fundamental Forces and Symmetries of the Universe''
with contract MRTN-CT-2004-005104,
the INTAS contract 03-51-6346 ``Strings, branes and higher-spin gauge fields'',
the Greek Ministry of Education programs $\rm \P Y\Th A\G OPA\S$ with contract 89194 and
the program $\rm E\Pi A N$ with code-number B.545.



\begin{thebibliography}{99}

\bibitem{coset}
  M.B.~Halpern,
  Phys. Rev. {\bf D4} (1971) 2398;\hfill\break
  P.~Goddard, A.~Kent and D.I.~Olive,
  Phys. Lett. {\bf B152} (1985) 88.


\bibitem{gwzw}
  K.~Bardakci, E.~Rabinovici and B.~Saering,
  Nucl. Phys. {\bf B299} (1988) 151;\hfill\break
  H.J.~Schnitzer,
  Nucl. Phys. {\bf B324} (1989) 412;\hfill\break
  D.~Karabali, Q.H.~Park, H.J.~Schnitzer and Z.~Yang,
  Phys. Lett. {\bf B216} (1989) 307.


\bibitem{IKOP2}
D. Isra\"el, C. Kounnas, D. Orlando and P.M. Petropoulos,
Fortsch. Phys. \textbf{53} (2005) 1030,  \texttt{hep-th/0412220}.

\bibitem{LS04}
D. Lowe and A. Strominger,
Phys. Rev. Lett. \textbf{73} (1994) 1468, \texttt{hep-th/9403186};
\hfill\break
C. Johnson,
Mod. Phys. Lett. \textbf{A10}, \texttt{hep-th/9409062};\hfill\break
D. Isra\"el, C. Kounnas, D. Orlando and P.M. Petropoulos,
Fortsch. Phys. \textbf{53} (2005) 73,  \texttt{hep-th/0405213}.

  \bibitem{LustCoset}
  D.~L\"ust,
  Nucl. Phys. {\bf B276} (1986) 220;\hfill\break
  L.~Castellani and D.~L\"ust,
  Nucl. Phys. {\bf B296} (1988) 143.


\bibitem{BN}
  I.~Bars and D.~Nemeschansky,
  Nucl. Phys. {\bf B348} (1991) 89.

\bibitem{wittenbh}
  E.~Witten,
  Phys. Rev. {\bf D44} (1991) 314.

\bibitem{BS1}
  I.~Bars and K.~Sfetsos,
  Mod. Phys. Lett. {\bf A7} (1992) 1091, {\tt hep-th/9110054}.

\bibitem{BS2}
  I.~Bars and K.~Sfetsos,
  Phys. Lett. {\bf B277} (1992) 269, {\tt hep-th/9111040}.

\bibitem{BS3}
  I.~Bars and K.~Sfetsos,
  Phys. Rev. {\bf D46} (1992) 4495, {\tt hep-th/9205037}.

\bibitem{BCR}
K.~Bardacki, M.J.~Crescimanno and E.~Rabinovici,
Nucl. Phys. {\bf B344} (1990) 344.

\bibitem{EFR}
  S.~Elitzur, A.~Forge and E.~Rabinovici,
  Nucl. Phys. {\bf B359} (1991) 581.


\bibitem{MSW}
  G.~Mandal, A.M.~Sengupta and S.R.~Wadia,
  Mod. Phys. Lett. {\bf A6} (1991) 1685.

\bibitem{BSexa}
  I.~Bars and K.~Sfetsos,
  Phys. Rev. {\bf D46} (1992) 4510, {\tt hep-th/9206006}.


\bibitem{DVV}
  R.~Dijkgraaf, H.L.~Verlinde and E.P.~Verlinde,
  Nucl. Phys. {\bf B371} (1992) 269.

\bibitem{haskins}
C.N. Haskins, Trans. Am. Math. Soc. {\bf 3}, (1902) 71.


\bibitem{weinberg}
S. Weinberg, {\it Gravitation and Cosmology}, John Wiley \& Sons, 1972.


\bibitem{qmparaf}
A.B. Zamolodchikov and V.A. Fateev,
  Sov. Phys. JETP {\bf 62} (1985) 215
  [Zh.\ Eksp.\ Teor.\ Fiz.\  {\bf 89} (1985) 380].

\bibitem{lykken}
  J.D.~Lykken,
  Nucl. Phys. {\bf B313} (1989) 473.



\bibitem{parafNON}
  K.~Bardakci, M.J.~Crescimanno and S.~Hotes,
  Nucl. Phys. {\bf B349} (1991) 439.

\bibitem{bardakciNONabel}
  K.~Bardakci,
  Nucl. Phys. {\bf B369} (1992) 461 and
  Nucl. Phys.  {\bf B391} (1993) 257.


\bibitem{bais}
  F.A.~Bais, P.~Bouwknegt, M.~Surridge and K.~Schoutens,
  Nucl. Phys. {\bf B304} (1988) 348.


\bibitem{PS05}
  P.M. Petropoulos and K. Sfetsos,
  JHEP \textbf{0601} (2006) 167, {\tt hep-th/0512251}.

\bibitem{sfet1}
  K.~Sfetsos,
  JHEP {\bf 9901} (1999) 015, {\tt hep-th/9811167}.

\bibitem{BakSfe}
  I.~Bakas and K.~Sfetsos,
  Phys. Rev. {\bf D54} (1996) 3995, {\tt hep-th/9604195}.

\bibitem{Lugo}
A.R.~Lugo,
  Phys. Rev. {\bf D52} (1995) 2266, {\tt hep-th/9411152} and
  Phys. Rev. {\bf D55} (1997) 6394, {\tt hep-th/9603182}.



\end{thebibliography}
\end{document}